\documentclass[10pt]{iopart}

\expandafter\let\csname equation*\endcsname\relax
\expandafter\let\csname endequation*\endcsname\relax

\usepackage{graphics}
\usepackage{physics}
\usepackage{xcolor}
\usepackage{enumerate}
\usepackage{amsmath}
\usepackage{amssymb}
\usepackage{graphicx}
\usepackage{float}

\usepackage{multirow}

\usepackage{hyperref}

\usepackage{iopams}

\begin{document}

\title[HFSS as a new tool for numerical plasma simulations]{ANSYS HFSS as a new numerical tool to study wave propagation inside anisotropic magnetized plasmas in the Ion Cylotron Range of Frequencies}

\author{V. Maquet$^1$, R. Ragona$^2$, D. Van Eester$^1$, J. Hillairet$^3$, F. Durodie$^1$}

\address{$^1$Laboratory for Plasma Physics, LPP-ERM/KMS, 1000 Brussels, Belgium\\
$^2$Technical University of Denmark, Department of Physics, 2800 Lyngby, Denmark\\
$^3$CEA, IRFM, F-13108 Saint-Paul-lez-Durance}
\ead{Vincent.Maquet@ulb.be}
\vspace{10pt}
\begin{indented}
\item[]Sometime in 2023
\end{indented}

\begin{abstract}
The paper demonstrates the possibility to use ANSYS HFSS as a versatile simulating tool for antennas facing inhomogeneous anisotropic magnetized plasmas in the Ion Cyclotron Range of Frequencies (ICRF). The methodology used throughout the paper is first illustrated with a uniform plasma case. We then extend this method to 1D plasma density profiles where we perform a first benchmark against the ANTITER II code. The possibility to include more complex phenomena relevant to the ICRF field in future works like the lower hybrid resonance, the edge propagation of slow waves, sheaths and ponderomotive forces is also discussed. We finally present a 3D case for WEST and compare the radiation resistance calculated by the code to the experimental data. 

The  main result of this paper -- the implementation of a cold plasma medium in HFSS -- is general and we hope it will also benefit to research fields besides controlled fusion. 
\end{abstract}

\vspace{2pc}
\\

%
%
%
\ioptwocol


\section{Introduction} \label{sec:Introduction}
Ion cyclotron Resonance Heating (ICRH) is one of the main heating schemes presently used in fusion devices and should represent a first-choice method in future fusion reactors like DEMO, ARC or CEFTR \cite{lin_wright_wukitch_2020,TRAN2022113159,Zhang_2022}. One of the main advantages of ICRH is its ability to directly heat ions inside the plasma core while not suffering any high density cutoff. Furthermore, ICRH covers a large range of applications besides its sole heating utility such as wall conditioning, plasma start-up, plasma control and plasma landings.

As ICRH antennas represent an opportunity for future fusion reactors, the accuracy of their design and modeling is essential to secure their correct operation. 
While a complete simulation of all physical aspects of an ICRH antenna is still beyond reach, numerical models developed towards this goal are constantly progressing \cite{Zhang_2022A} and any new simulation tool development represents an opportunity to further characterize those physical phenomena. Those aspects are presently regarded as essential in the ICRH community and studied in detail to correctly account for plasma wall interactions arising during ICRH operation such as ponderomotive effects and sheaths. 

Several simulating tools are already available in the Ion Cyclotron Range of Frequencies (ICRF) to include a realistic antenna geometry, edge plasma and sometimes account for the full tokamak geometry. A non-exhaustive list is given below:
\begin{itemize}
    \item TOPICA \cite{Lancellotti_2006,Milanesio_2009}, based on integral equations discretized via the method of moments, can include complex 3D antenna geometries. The antenna is placed in a cavity (port) and face a 1D hot plasma.
    \item REPLICASOL \cite{Tierens_2019}, based on the COMSOL \cite{COMSOL} finite element software, can include complex 3D antenna geometries. The antenna is placed in front of a 1D or 3D cold plasma profile with Perfect Matching Layers (PMLs) imposed far from the antenna.
    \item PETRA-M \cite{shiraiwa2017rf,Bertelli2020}, based on the MFEM finite element discretization library \cite{kolev2010mfem}, can include complex 3D antenna geometries. The antenna is placed in front of a 1D or 3D plasma profile with Absorbing Boundary Conditions (ABL) far from the antenna. The code was also used to simulate full 3D tokamak cold plasma and includes the possibility to simulate RF sheaths \cite{Shiraiwa_2023}. The code also includes the possibility to couple to TORIC.
    \item VSIM \cite{Smithe2007}, based on a time domain approach, can include complex 3D antenna geometries and allows to simulate the full tokamak geometry. The code can also include nonlinear sheath BCs.
\end{itemize} 
This paper aims at adding the widely used ANSYS HFSS \cite{ANSYS} software to this restrained list of codes. The possibility to include a plasma tensor in HFSS was already demonstrated in the lower hybrid range of frequencies for \textit{diagonal} but anisotropic and inhomogeneous cold plasma dielectric tensors \cite{HILLAIRET20191473} and benchmarked using the open-source 1D code ALOHA \cite{Hillairet_2010}. The aim of this paper is to extend this methodology to the ICRF, characterized by the presence of non-negligible off-diagonal terms in the plasma dielectric tensor and to explore the potential and the limitations of such an approach.
\newline

The paper is organized as follows.
Section \ref{sec:uniform} describes the implementation of a uniform cold plasma medium.
Some key properties expected in a cold plasma are tested and the possibility to implement a rotation of the dielectric tensor is verified. Section \ref{sec:profile} extends the approach described in section \ref{sec:uniform} to plasma profiles whitin HFSS. 
The possibility to achieve high single pass absorption with an absorption layer imposed at the end of the domain is first discussed. 
A first benchmark of the simulation performed in HFSS is then executed using the 1D ANTITER code \cite{messiaen2010performance}. 
The possibility to include the slow wave and the lower hybrid resonance inside those profiles is explored in the limit of high collisionality. A correction of the cold plasma medium to compute the power deposition on each plasma species in post-processing is presented. The possibility to account for ponderomotive forces and sheath are finally discussed. 
Section \ref{sec:3D} discuss a second benchmark of the code. The radiation resistance and the voltages at the feeders is calculated by implementing a 3D plasma model based on a WEST discharge. It is then compared against the resistance and voltages measured form the experimental data.
\newline

It is important to note that, while this paper concentrates on the simulation of waves propagating inside a cold plasma in the ICRF, the methodology proposed to implement the cold plasma dielectric tensor is general and should work in any plasma research field. We therefore hope that this paper will benefit to a larger plasma scientific community (space plasmas, plasma etching, plasma propulsion, etc.).

\section{Uniform Plasma} \label{sec:uniform}
The possibility to include off-diagonal terms in the dielectric tensor in HFSS is first explored using a simple uniform cold plasma case in the ICRF. 
\newline

The cold plasma dielectric tensor is conventionally expressed in a coordinate system $(\vb{e}_{\perp,1},\vb{e}_{\perp,2},\vb{e}_{\parallel})$ where $\vb{e}_{\parallel}$ is aligned to the total magnetic field direction:
\begin{align}
    \epsilon \equiv \mqty[\epsilon_1 & i\epsilon_2 & 0 \\ -i\epsilon_2 & \epsilon_1 & 0 \\ 0 & 0 & \epsilon_3] \equiv \mqty[S & -iD & 0 \\ iD & S & 0 \\ 0 & 0 & P],
    \label{eqICRH:coldPlasmaTensor}
\end{align}
with 
\begin{align}
    \begin{split}
        \epsilon_1 \equiv S &\equiv 1 - \sum_s \frac{\omega_{ps}^2}{\omega^2 - \omega_{cs}^2}, \\
        -\epsilon_2 \equiv D &\equiv \sum_s \frac{\omega_{cs}}{\omega}\frac{\omega_{ps}^2}{\omega^2 - \omega_{cs}^2}, \\
        \epsilon_3 \equiv P &\equiv 1 - \sum_s \frac{\omega_{ps}^2}{\omega^2},
    \end{split}
\end{align}
where $S$, $P$ and $D$ are the so-called Stix parameters \cite{stix1992waves,swanson2012plasma}. $S$ and $D$ terms stand for the half-sum and the half-difference of left ($L$) and right ($R$) terms.

In the ICRF, two distinct set of waves can propagate inside the plasma: the fast and the slow magnetosonic waves. In the cold plasma limit, the dispersion relation of each wave can be approximated by: 
\begin{align}
    k_{\perp,\text{FW}}^2&\approx \frac{(k_0^2 R-k_\parallel^2)(k_0^2L-k_\parallel^2)}{k_0^2\epsilon_1-k_\parallel^2}, \label{eqICRH:kperpF}\\
    k_{\perp,\text{SW}}^2&\approx\frac{\epsilon_3}{\epsilon_1} (k_0^2\epsilon_1-k_\parallel^2). \label{eqICRH:kperpS}
\end{align} 

To include losses inside the cold plasma medium, one can add a small positive imaginary part in the plasma dielectric tensor. This correction can be introduced based on a simple Krook model of the collision term $\nu$ entering Boltzmann equations \cite{swanson2012plasma} like
\begin{align}
    \begin{split}
        \epsilon_1 \equiv S &\equiv 1 - \sum_s \frac{\omega+i\nu_j}{\omega}\frac{\omega_{ps}^2}{(\omega+i\nu_j)^2 - \omega_{cs}^2}, \\
        -\epsilon_2 \equiv D &\equiv \sum_s \frac{\omega_{cs}}{\omega}\frac{\omega_{ps}^2}{(\omega+i\nu_j)^2 - \omega_{cs}^2}, \\
        \epsilon_3 \equiv P &\equiv 1 - \sum_s \frac{\omega_{ps}^2}{\omega(\omega+i\nu_j)}.
    \end{split}
    \label{eqICRH:coldPlasmaTensorcorr}
\end{align}
The collision term is of the order of $10$ kHz inside the plasma core. The ratio $\nu/\omega$ is thus of the order of $10^{-4}$ \cite{koch2008coupling}, and the dissipative effects due to collisions is very small. In the edge, the plasma is colder and collisions with neutrals may be dominant, potentially leading to higher ratios $\nu/\omega$. Using this collision ratio inside the tensor expressions leads to small imaginary corrections of each tensor terms. In this paper, the  average collision ratio $\nu/\omega$ was fixed to $10^{-3}$, if not specified otherwise\footnote{Taking $\abs{S}$ and $\abs{D}$ order of $10^{3}$ and $\abs{P}$ of the order of $10^6$, this leads to a correction of the order unity in $S$ and $D$ and of the order of $10^3$ for $P$.} and, provided this value is not close to unity,does not affect the results presented.
\newline

Directly defining the dielectric tensor in HFSS lead to numerical problems. To succesfully implement a plasma medium in HFSS the conductivity tensor $\sigma$ definition was used instead of the dielectric one:
\begin{align}
    \epsilon &= I + \frac{i}{\epsilon_0\omega} \sigma.
\end{align}
To be noted that the convention used in our work is $(-i\omega t)$ while the HFSS convention uses $(+i\omega t)$. The above procedure essentially enables to overcome numerical instabilities encountered in HFSS due to the large values of the Stix $P$ component of the cold plasma dielectric tensor. While it has not been tested yet, this trick could also be used for other software like COMSOL. 

\subsection{Plane wave in a uniform plasma} \label{subsec:Plane_wave}
The propagation of an incident plane wave at 55 MHz inside a uniform plasma was used to test the validity of the approach described in the previous paragraph. For the rest of the manuscript, the background magnetic field direction is chosen to be aligned with the toroidal direction such that $\vb{e}_{\parallel}\equiv\vu{z}$, unless specified otherwise. Uniform cold plasma Stix components $S\approx-500$, $D\approx-10^3$ and $P\approx-10^6$ are chosen such that the slow wave is evanescent and the fast wave is propagative. Those values are close to values expected near the core of a fusion plasma. With these parameters, using \eqref{eqICRH:kperpF}, the wavelength of the fast wave is $\lambda\approx0.14$ m.
\newline

The boundaries of the plasma domain are periodic in the toroidal $\vu{z}$ and poloidal $\vu{y}$ directions. In the radial $\vu{x}$ direction, exponential losses are introduced in the diagonal terms of the conductivity tensor as an absorbing boundary region in order to damp the wave far from the region it was initially launched. The loss value is chosen constant near the antenna\footnote{Of the order of $10^{-3}$, corresponding to the imaginary part present in the dielectric tensor of the order of unity.} and grows exponentially after 0.5 m. In this example, the absorbing boundary layer chosen is not optimal and can create non-negligible nonphysical reflections at the boundary of the domain. The possible optimization of such an absorbing layer is discussed in the next section.
\newline

\begin{figure*}[tb]
    \centering
    \includegraphics[width=0.7\linewidth]{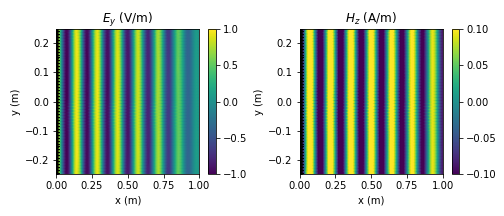}
    \caption{$E_y$ and $H_z$ waves characteristic of the FW propagation in the medium.}
    \label{fig:planewave}
\end{figure*}

The outcome of the simulation is presented in figure \ref{fig:planewave}, showing the $E_z$ and $H_y$ fields of the fast wave propagating radially inside the plasma. Measuring the distance between two field maxima, we find a wavelength $\lambda\approx0.14$ m as expected from equation \eqref{eqICRH:kperpF}. One can also see that the waves are not fully absorbed at the end of the media showing the necessity to further optimize the absorbing layer at the end of the domain. Fortunately, this lack of absorption does not seem to influence the wavelength measured in our simulation.

\subsection{Antenna excitation inside a uniform plasma} \label{subsec:small_antenna}
Using the same plasma parameters, a simple ICRH antenna box made of one radiating strap was also tested. The antenna characteristics are summarized in table \ref{tab:antenna_ref}. They are deliberately close to the strap characteristics of the WEST IC antennas and will be used for most of the examples presented in the manuscript. The driving frequency of the antenna is chosen to be 55 MHz. The antenna box is defined as vacuum and the antenna aperture is placed in front of the plasma. The antenna does not include a Faraday Shield (FS) as the plasma is playing the role of the polarizer, preventing the slow wave from propagating.

\begin{table}[h]
    \centering
    \begin{tabular}{c|c}
       Strap width & 130 mm \\
       Strap height & 280 mm \\
       Strap thickness & 15 mm \\
       Strap recess & 40 mm \\
       Box depth & 200 mm \\
       Box width & 200 mm 
    \end{tabular}
    \caption{Simplified antenna dimensions used for the manuscript.}
    \label{tab:antenna_ref}
\end{table}

\begin{figure*}[tb]
    \centering
    \includegraphics[width=0.45\linewidth]{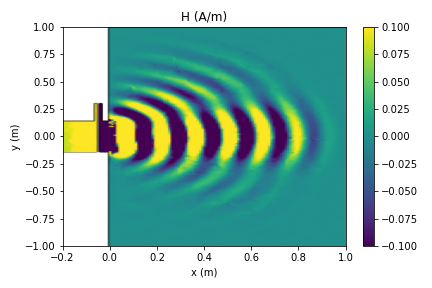}
    \includegraphics[width=0.45\linewidth]{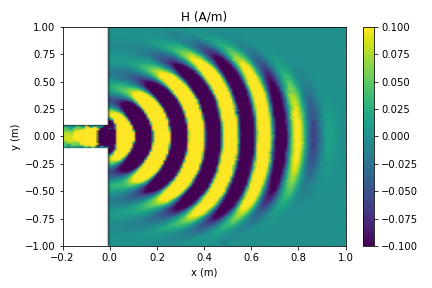}
    \caption{(a) Poloidal and (b) toroidal cut of HFSS H fields in the uniform cold plasma defined. The fast wave wavelength expected using \eqref{eqICRH:kperpF} is retrieved.}
    \label{fig:Hfields_uniform}
\end{figure*}

The toroidal and poloidal magnetic field excited by the antenna are presented in figure \ref{fig:Hfields_uniform}. We find again a main wavelength $\lambda\approx0.14$ m excited by the antenna and propagating away from it. This simulation can run in less than three minutes on a 32 GB RAM machine despite the large plasma volume considered.

\subsection{Plasma rotations} \label{subsec:Rotation}
The simple uniform plasma described in the previous paragraph is now rotated by an angle $\alpha$. In ICRH physics, a rotation of the magnetic field is often used to take into account the poloidal component of the magnetic field present in tokamaks. In our case, the tilt of the magnetic field $\alpha$ is performed around the radial direction $\vb{e}_x$. This rotation can either be directly implemented in HFSS through a change of reference frame definition, where the plasma dielectric tensor is defined, or inside the conductivity tensor defined in Python and imposed in HFSS. The rotated tensor $\epsilon'$ can be expressed in terms of the initial cold plasma dielectric tensor and the rotation matrix $R(\alpha)$ as
\begin{align}
    \epsilon' = R(\alpha)^T \epsilon \, R(\alpha),
\end{align}
where $\alpha$ is the imposed rotation angle.

\begin{figure*}
    \centering
    \includegraphics[width=0.49\linewidth]{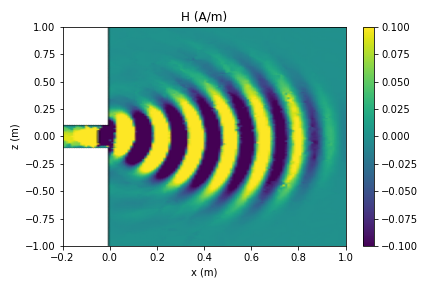}
    \includegraphics[width=0.49\linewidth]{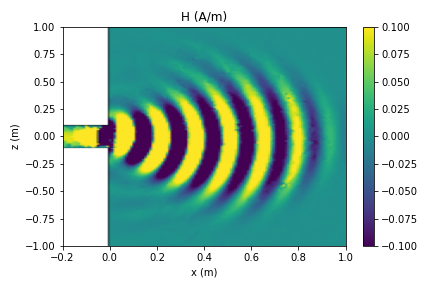}
    \caption{Toroidal cut of H field obtained in HFSS using the cold uniform plasma definition with a rotation of 30$^\circ$ (a) performed inside the conductivity tensor, (b) performed inside HFSS}
    \label{fig:Hfields_uniform_rot}
\end{figure*}

An angle of $\alpha=30^\circ$ is imposed in order to clearly see the effect of such a tilt on the fast wave propagation. The goal is to compare the rotation performed inside HFSS against the rotation performed inside the dielectric tensor. In figure \ref{fig:Hfields_uniform_rot} one finds the same effect on the fields. It should therefore be possible to implement a rotation for more complex plasma shapes, and taking into account the toroidal and poloidal rotations present inside a tokamak or a stellarator.

\section{Plasma Profiles} \label{sec:profile}
Now that a uniform cold plasma inside HFSS was successfully implemented, the possibility to impose a cold plasma \textit{profile} is explored. Including a 1D profile in HFSS is straightforward and documented. Therefore, the goal of this section is to first optimize and benchmark the plasma simulation performed in HFSS with the semi-analytical code ANTITER. The possibility to characterize critical ICRH aspects like the lower hybrid resonance, the slow wave propagation, sheath and the ponderomotive forces is discussed.

\subsection{Absorbing Boundary Layer optimization} \label{subsec:ABL}
The results shown in section \ref{sec:uniform} were performed with an absorbing boundary layer (ABL)\footnote{Also known as adiabatic absorber.} which relies on the introduction of losses analytically inside the HFSS diagonal conductivity terms of our cold plasma dielectric tensor formulation. This method introduces nonphysical reflections in the simulations. Furthermore, the amount of those reflections is potentially important since the exponential loss factor used was not optimized and will be quantified in the next paragraphs.
\newline

The best solution to avoid those reflection would be to implement Perfectly Matching Layers (PMLs) at the boundaries of the plasma domain \cite{Jacquot_2013,Louche2017}. However, the implementation of PMLs is either complex or just incompatible with our plasma formulation in HFSS. Indeed, the coordinate stretching functions used in PMLs does not only apply on the conductivity tensor alone but also on the dielectric and permittivity identity matrices present in Maxwell's formulation (see equation, (6a)-(6d) in \cite{Jacquot_2013}). Implementing this coordinate stretching inside the dielectric tensor would lead back to the numerical problems faced prior the conductivity formulation presented in the first section. Furthermore, PMLs can only damp one of the two waves propagating into our plasma, which can become problematic for the following discussions on the slow wave propagation and absorption. For these reasons, PMLs implementation was not pursued in the present work and could be the subject of future analysis.
\newline

Instead of PMLs, ANTITER\footnote{ANTITER describes an antenna facing a plasma in plane geometry. The code assumes infinitely thin straps and an ideal Faraday shield at the box aperture.} was used to optimize an absorbing boundary layer. An analytical profile was used for the ratio $\nu/\omega$ entering in \eqref{eqICRH:coldPlasmaTensorcorr} in ANTITER. The difference between the output FW admittance matrix obtained with this profile and the analytical admittance matrix of the FW radiation condition $\xi_b$ used in ANTITER was minimized. In accordance with the absorbing layer theory \cite{oskooi2008failure}, shorter ABL depth leads to higher differences between theoretical and numerical admittance matrices computed in ANTITER. This minimization was performed over a parameter $a$ for several analytical functions $f(x)$ and for a fixed boundary layer length $L_{ABL}$ of $0.5$ m following
\begin{align}
    \sigma/\omega = \left\{
    \begin{array}{cc}
        0 & \text{for} \quad 0<x<0.5,  \\
        a f(x) & \text{for} \quad 0.5<x<1.
    \end{array}
    \right.
\end{align} 
While the exact value of $a$ can change from one plasma to another, the best function $f(x)$ found to minimize the reflection of the FW at the boundary for a large number of wavelength $k_y$ and $k_z$ was of the form similar to a $n$-order PML stretching function
\begin{align}
    f(x) = \qty(\frac{x-L_{r}}{L_\text{ABL}})^n.
\end{align}
In our formulation, the correction made to the conductivity tensor is not only applied on the diagonal terms of the tensor but also on its off-diagonal terms. 
\newline

From theory, one can compute the standing wave ratio (SWR) and the reflection $\Gamma$ coefficient of the ABL provided the presence of a minimum and a maximum of the electric field in the unperturbed radial domain following
\begin{align}
    SWR &= \frac{\max(\abs{E})}{\min(\abs{E})}, \\
    \Gamma &= \frac{\text{SWR}-1}{\text{SWR}+1}.
\end{align}
Using the plane wave excitation example used in section \ref{subsec:Plane_wave}, one can extract the $E_y$ electric field component of the fast wave and compute the reflection induced by the ABL of our model. Table \ref{tab:reflections} summarize different ABLs and their reflection in the uniform plasma described in section \ref{subsec:Plane_wave}. The first two entries of the table verify the perfect reflection obtained for the metallic boundary of HFSS and shows that the radiating boundary function of HFSS is not suited for our problem. This was expected as the background material of the radiating boundary is different than the material in the calculation domain. The third entry of table \ref{tab:antenna_ref} demonstrates that the ABL used in the uniform cases of section \ref{sec:uniform} was non-optimal and led to high reflections. The rest of the table shows the improvement obtained when using the new approach presented in this section.

ANTITER computations predict a minimum of reflection for PML-like function $f(x)$ of the 3rd order with a loss factor $a\approx2$. ANTITER also predicts a minimum of reflection for the $f(x) = e^\qty(1 - L/x)$ with a loss factor of $a=0.8$, with a reflection coefficient higher than the PML-like functions. These predictions are verified numerically with HFSS in Table \ref{tab:reflections}. However, ANTITER predicts a broad minimum for a fourth order PML-like function $f(x)$ with a loss factor $a\approx2.2$ which is not verified in HFSS.

\begin{table}[]
\centering
\begin{tabular}{c|c|c}
$f(x)$ & $a$ & $\Gamma$ for $E_y$\\ \hline
\multicolumn{2}{l|}{HFSS metallic boundary} &   $>0.99$\\
\multicolumn{2}{l|}{HFSS radiating condition} &  0.98\\
\multicolumn{2}{l|}{ABL defined in section \ref{sec:uniform}} &  0.2\\ \hline
\multicolumn{1}{l|}{\multirow{3}{*}{$\qty(\frac{x-L_{r}}{L_\text{ABL}})^3$}} & 1    &  0.0040 \\
\multicolumn{1}{l|}{}                  &   2  & 0.0005 \\
\multicolumn{1}{l|}{}                  &   3  & 0.0010 \\ \hline
\multicolumn{1}{l|}{\multirow{3}{*}{$\qty(\frac{x-L_{r}}{L_\text{ABL}})^4$}} & 1    &  0.01 \\
\multicolumn{1}{l|}{}                  &   2  & 0.0007 \\
\multicolumn{1}{l|}{}                  &   3  & 0.0004 \\
\multicolumn{1}{l|}{}                  &   4  & 0.0002 \\
\multicolumn{1}{l|}{}                  &   5  & 0.0005 \\ \hline
\multicolumn{1}{l|}{\multirow{3}{*}{$e^\qty(1 - L/x)$}} & 0.3    &  0.043 \\
\multicolumn{1}{l|}{}                  &   0.8  & 0.0011 \\
\multicolumn{1}{l|}{}                  &   1.7  & 0.0019 \\
\end{tabular}
\caption{Reflection for different Absorbing Boundary Layer functions $f(x)$ and factors $a$ in HFSS.}
\label{tab:reflections}
\end{table}

While ANTITER exact minimum prediction capabilities are not optimal, reflections computed in HFSS gives us confidence that, following ANTITER outputs, one can greatly reduce the reflections obtained at the end of the plasma domain. For the rest of the text, 3rd order PML-like loss functions are used with a loss factor minimizing the reflections of most of the wavenumbers $k_y$ and $k_z$ excited by ICRH antennas in ANTITER. This should guarantee a reflection coefficient of $\Gamma\leq0.01$.

\subsection{ANTITER comparison} \label{subsec:ANTcomparison}
\begin{figure}[h]
    \centering
    \includegraphics[width=0.95\linewidth]{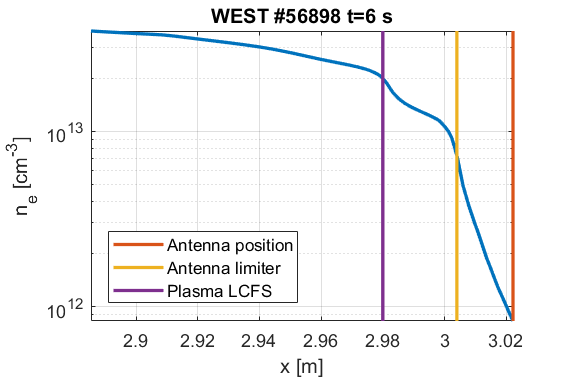}
    \caption{Electron density profile WEST shot \#56898 at 6 s along with the antenna front face, the antenna limiter and the LCFS positions.}
    \label{fig:dens_profile}
\end{figure}

Now a plasma profile is imposed at the beginning of the domain on top of the ABL profile imposed at the end of it. A 1D density profile is considered with parameters coming from the WEST pulse \#56898. In the experiment, the plasma composition is D-(H) with a minority concentration around 4\%. The electron density profile is reconstructed from reflectometry data and shown in figure \ref{fig:dens_profile} for $t=6\text{ s}$. The reflectometer is toroidally located between the Q1 and Q4 WEST ICRF antennas, on the equatorial plane of the machine.

In the chosen profile, the fast wave can propagate and the slow wave is strongly evanescent. For this specific reason, the slow wave will not couple to the plasma load and the addition of a Faraday shield should not drastically change the loading seen by the antenna. We used the absorbing boundary layer described in section \ref{subsec:ABL}, and compared the antenna response with the one computed in ANTITER.

\subsubsection{Antenna impedance comparison}\label{subsubsect:Antenna_impedance} \

For our first comparison with ANTITER, we will compare the output impedance of the one strap antenna introduced previously. 
\newline

For this comparison to be relevant, one needs to get as close to the ANTITER assumptions as possible. In HFSS the ideal Faraday shield function is carried out by the plasma at the aperture of the antenna box. The strap thickness of the antenna as a function of the output impedance measured in HFSS is presented in table \ref{tab:antenna_ref}. The strap thickness was reduced from 20 to 2.5 mm. As the strap thickness is decreased, we get closer to the impedance $Z_{strap}$ predicted by ANTITER despite the simplicity of the ANTITER code antenna geometry description. 

\begin{table}[h]
    \centering
    \begin{tabular}{c|c}
       Strap &  $Z_{\text{mat}}$ ($\Omega$) \\ thickness & with ABL \\ \hline 
       15 mm & 1.27 + 22.25i \\
       10 mm & 1.38 + 23.14i \\
       5 mm & 1.53 + 24.17i \\
       2.5 mm & 1.6 + 24.8i \\
       ANTITER & 1.68 + 23.56i \\
    \end{tabular}
    \caption{Impedance of the strap $Z_{strap}$ found with HFSS and ANTITER.}
    \label{tab:Zmat_HFSS}
\end{table}

\subsubsection{Comparison with Q2 antenna} \

The power spectrum and the fields of a flat simplified Q2 antenna geometry compatible with ANTITER are now compared, for an excitation of 1 A at the straps. Although not shown as it was already performed in section \ref{subsubsect:Antenna_impedance}, the impedance matrices found between ANTITER and HFSS are very close.
\newline

\begin{figure*}
    \centering
    \includegraphics[width=\linewidth]{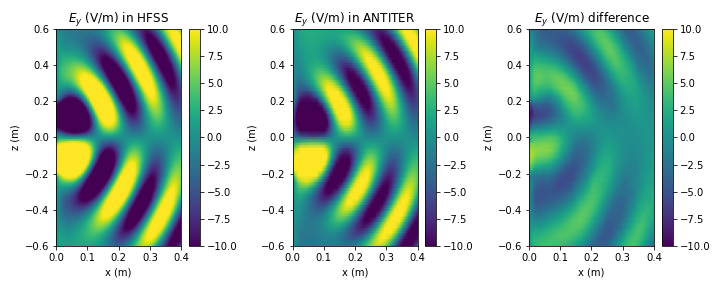}
    \includegraphics[width=\linewidth]{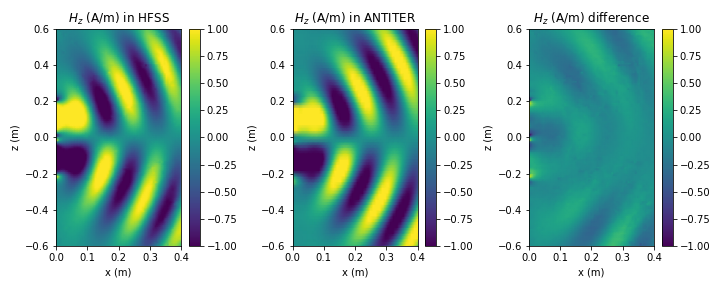}
    \caption{Comparison of the (top) $E_y$ and (bottom) $H_z$ field found in HFSS and in ANTITER as well as their difference.}
    \label{fig:ANTvsHFSS_Ez&Hz}
\end{figure*}

The $E_y$ and $H_z$ amplitude field maps found in HFSS and ANTITER as well as the difference found between them is presented in figure \ref{fig:ANTvsHFSS_Ez&Hz}. One can see that the field amplitude and distribution found in ANTITER is retrieved in HFSS. 

\begin{figure*}
    \centering
    \includegraphics[width=\linewidth]{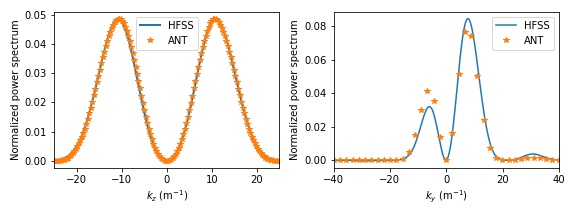}
    \caption{Comparison of the (left) toroidal and (right) poloidal power spectra calculated from HFSS and ANTITER.}
    \label{fig:ANTvsHFSS_powspctr}
\end{figure*}

The comparison of the normalized toroidal and poloidal power spectra found in HFSS is presented in figure \ref{fig:ANTvsHFSS_powspctr}. The power spectrum calculated from HFSS is in excellent agreement with the one calculated by ANTITER. The small discrepancy in the poloidal spectrum can be explained by the fact that ANTITER is assuming a constant current on the strap while HFSS does calculate the current self-consistently. As the amplitudes of the waves are similar, one would also retrieve similar non-normalized toroidal and poloidal spectra.

\subsection{The slow wave and the lower hybrid resonance} \label{subsec:SW}
After the 1D benchmark presented, one can explore the possibility to characterize the slow wave propagation, and its absorption at the lower hybrid (LH) resonance, a critical aspect of ICRH physics.
\newline

In the ICRF, a LH resonance can appear in the edge of a tokamak discharge when the antenna aperture is facing a low plasma density (usually below $10^{17}$ m$^{-3}$).
This scenario is expected to take place in future devices with large scrape-off layers, like ITER, CEFTR and DEMO. 
This resonance can create deleterious edge losses and can lead to the presence of non-desired modes in between the wall of the device and the LH resonance \cite{Messiaen_2020, maquet_messiaen_2020}. 
At these low densities, the slow wave is propagative and can either be excited 
\begin{itemize}
    \item by a confluence of the fast wave with the slow wave for wavelength $\abs{k_\parallel}<k_0$, where $k_\parallel$ and $k_0$ are respectively the parallel and the vacuum wavenumbers\cite{Messiaen_2020};
    \item by direct excitation of the slow wave due to complex boundaries \cite{kohno2015numerical} or to a misalignment of the antenna Faraday shield with the total confining magnetic field \cite{Messiaen_2021}.
\end{itemize} 
The possibility to include the LH resonance in HFSS could be very useful to study the effect of this resonance on the electromagnetic fields near complex ICRH antenna structures, like the limiters or the FS geometry, and could be of importance for RF sheath calculations. 
Therefore, implementing the slow wave in HFSS would help to lean a step toward realistic simulation of complex 3D systems and their interaction with the edge.
However, the lower hybrid resonance and the slow wave are characterized by small wavelength waves known to be numerically challenging to simulate.
\newline

\begin{figure}
    \centering
    \includegraphics[width=\linewidth]{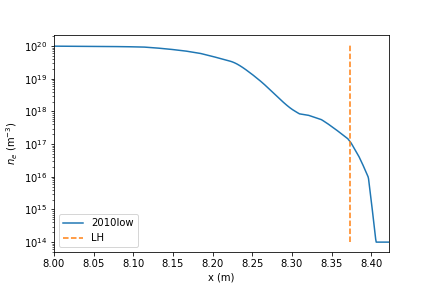}
    \caption{ITER 2010low profile electron density profile. The LH resonance position is made visible.}
    \label{fig:ITER2010low}
\end{figure}

The possibility to include such physics in HFSS is first explored by means of the single strap antenna model, facing one of the reference density profile of ITER (2010low -- \cite{CARPENTIER2011S165}). 
This profile is presented in figure \ref{fig:ITER2010low} along with the LH resonance position. 
The LH resonance is located approximately 50 mm in front of the antenna. 
The magnetic field, minor and major radii used are respectively $B_0=5.3$ T, $R_0=6.2$ m, $a=2$ m. The antenna radial position is $x=8.422$ m. 
The wave frequency is set to $55$ MHz and the plasma composition chosen is a $0.56$-$0.44$ deuterium-tritium mix. Periodic boundary conditions are imposed in the toroidal and poloidal directions of the plasma, and an optimized ABL is imposed in the radial direction, opposite from the antenna.
\newline

With these plasma parameters and the ratio $\nu/\omega \approx 10^{-3}$ previously defined, equation \eqref{eqICRH:kperpS} predicts a slow wave wavelength between $15$ to $0.2$ mm near the LH resonance, for toroidal wavenumbers $k_z$ between $0.5$ and $25$ m$^{-1}$. The grid size requirements to resolve such small wavelength is beyond the capabilities of our computing resources. However, by using a smaller one-strap model with larger fictitious losses $\nu/\omega = 0.5$, we were able to retrieve the expected physical characteristics of a slow wave that propagates towards the LH resonance.

\begin{figure*}
    \centering
    \includegraphics[width=\linewidth]{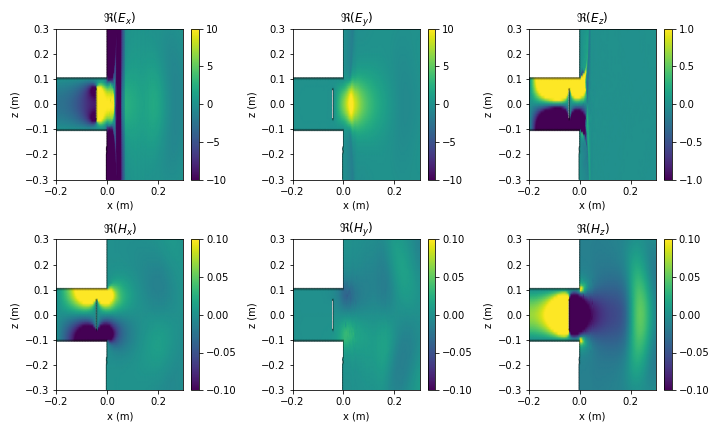}
    \caption{$\Re(E_x)$, $\Re(E_z)$ and $\Re(E_z)$  field close to the antenna launcher. $\Re(E_z)$ fields aligning to the background magnetic field direction characteristic of the slow wave are recognizable. Large $\Re(E_x)$ fields at the LH resonance and constant fields between the wall and the LH resonance characteristic of the coaxial modes are also visible. The beginning of the fast wave excited at the antenna is visible through $\Re(H_z)$.}
    \label{fig:SWEfields}
\end{figure*}

The first characteristic expected is the alignment of the slow wave to the background magnetic field while approaching the LH resonance. This is verified in figure \ref{fig:SWEfields} where the $E_z$ component of the electric field, characteristic of the slow wave polarization, is aligning to the $\vu{z}$ direction of the background magnetic field. The presence of a strong $E_x$ field amplitude at the LH resonance and a large $E_x$ field amplitude between the wall and the LH resonance, characteristics of propagating coaxial modes \cite{Messiaen_2020}, are visible in figure \ref{fig:SWEfields}, and were also expected.
\newline

Those first results show that HFSS is capable of capturing the effects of the LH resonance. Some of the numerical limitations we have encountered during this study will be overcome by expanding the computing resources at our disposal for the forthcoming studies.

\subsection{Tepid dielectric tensor} \label{subsec:Tepid}
Accounting for kinetic effects adds finite Larmor radius (FLR) - \textit{i.e} temperature - corrections to the dielectric tensor elements. Assuming the electric field phase change that a charged particle witnesses when orbiting around the local magnetic field line is modest, FLR corrections are conventionally introduced by making a Taylor series expansion of the dielectric response in terms of $k_\perp \rho_L$, where $k_\perp$ is the perpendicular wave number and $\rho_L$ is the Larmor radius. The dielectric tensor is typically referred as "hot´´ when the series expansion is truncated at second order in $k_\perp \rho_L$ (FLR$_2$), while retaining all corrections is commonly labeled as "full hot" (FLR$_\infty$). 

One can introduce the lowest order FLR$_0$ correction to the cold plasma expression \eqref{eqICRH:coldPlasmaTensor} -- hereafter referred as tepid -- following this line of though:
\begin{align}
    R_H &= 1 + \sum_s \frac{\omega_p^2}{\omega^2} \zeta_0 Z(\zeta_{-1}), \\
    L_H &= 1 + \sum_s \frac{\omega_p^2}{\omega^2} \zeta_0 Z(\zeta_{+1}), \\
    P_H &= 1 + \sum_s \frac{\omega_p^2}{\omega^2} 2\zeta_0^2 \qty[1 + \zeta_0 Z(\zeta_0)], 
\end{align}
where $\zeta_N = (\omega - N\Omega)/(k_\parallel v_{th} \sqrt{2})$. Note that these expressions now depend on the parallel wave number $k_\parallel$. Rigorously accounting for parallel gradients of the electric field requires a more involved procedure than what is intended in the present work. In practice, we will assume that poloidal field corrections to the parallel wave number are modest, so that they can be neglected as a first approximation. Adopting FLR$_0$ and making this supplementary assumption allows to generalize the dielectric tensor without needing major modifications to the model: it amounts to replacing the cold plasma terms by their finite temperature expressions but avoids the complication of the dielectric tensor becoming a differential, or integro-differential, operator rather than a matrix.

By selecting the main parallel wave number excited by a given antenna power spectrum, FLR$_0$ gives a first approximation of the power that will be absorbed inside the plasma core. This simple correction allows accounting not only the edge but also the core of the plasma domain into consideration. However, this correction does not account for finite Larmor effects and thus cannot take into account harmonic heating.
\newline

In the post-processing phase, one can approximate the total absorbed power and decompose the contribution of each species through the relations
\begin{align}
    P_{abs,s} = \frac{1}{2} \Re{\vb{E}^* \vdot \vb{J}_s},
    \label{eq:Pabs_tepid}
\end{align}
where 
\begin{align}
    \vb{J}_s = \frac{k_0^2}{i\omega\mu_0} (\epsilon-I)\vdot \vb{E}.
\end{align}

\begin{figure*}
    \centering\includegraphics[width=\linewidth]{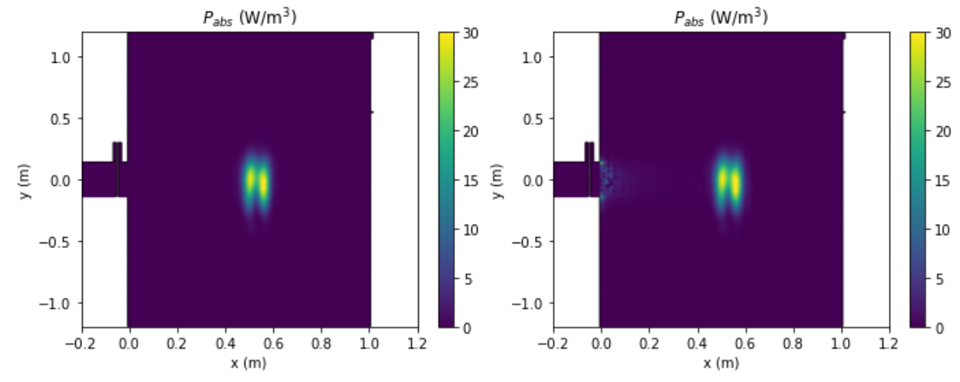}
    \caption{Power loss density on a plane compute in (left) Python and (right) HFSS. The two agree provided the introduction of a factor 2 in HFSS that could not be explained as the definition used should be the same.}
    \label{fig:comparison_Pabs}
\end{figure*}

To verify that procedure, the former one strap example is extended to include the resonance present in the plasma core. Periodic boundary conditions are imposed in the poloidal and toroidal directions far from the radiating antenna and the radial boundary is a metallic wall located at 1 m from the antenna. For verification we first compute the volume power loss computed in HFSS and compare it with our volume loss $P_{abs,tot}$. This comparison is presented in figure \ref{fig:comparison_Pabs}. Apart from a factor 2\footnote{Factor that we could not explain as the definition used is strictly the same as the one described in HFSS documentation and that the fields used are the one extracted from HFSS.}, the volume loss computed inside HFSS and with the relation \ref{eq:Pabs_tepid} are identical.
\newline

\begin{figure*}
    \centering\includegraphics[width=\linewidth]{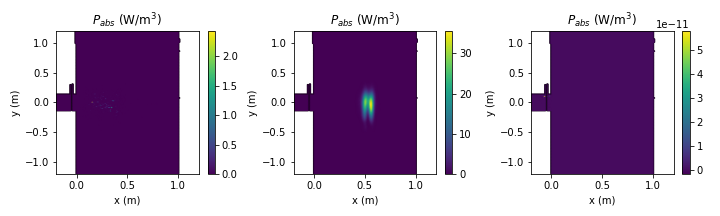}
    \caption{Power deposition on (left) electrons, (middle) Hydrogen and (right) deuterium. From theory we expect low deposition on the electrons, a large deposition on the hydrogen minority species and a no deposition on the deuterium majority due to the incapacity of the correction introduced to capture $N>1$ harmonic heating.}
    \label{fig:power_deposition}
\end{figure*}

One can then compute the power deposited on each species composing the plasma. From theory, with minority heating we expect most of the power to be deposited on the minority species H. We can expect a small fraction of electron heating and, due to the intrinsic assumptions of the correction used, we do not expect any heating of D. This is verified in figure \ref{fig:power_deposition}.

\begin{figure*}
    \centering
    \begin{minipage}{0.5\linewidth}
        \centering
        \includegraphics[width=\linewidth]{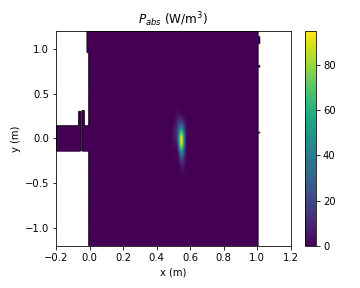}
    \end{minipage}%
    \begin{minipage}{0.5\linewidth}
        \centering
        \includegraphics[width=\linewidth]{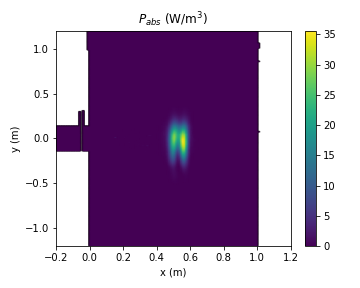}
    \end{minipage}
    \caption{Power deposition for (left) $k_\parallel=5$ m$^{-1}$ and (right) $k_\parallel=10$ m$^{-1}$. As expected form theory, a smaller width is found of lower $k_\parallel$.}
    \label{fig:kz5vskz10}
\end{figure*}

Finally, at the cyclotron resonance position $R_{ci}$, one can also verify that the resonance width $\delta R$ is proportional to the parallel wavenumber $k_\parallel$ following $\delta R/R_{ci}\propto k_\parallel v/\omega$. This is verified in figure \ref{fig:kz5vskz10} for two $k_\parallel$ values.
\newline

A next step could be to implement the kinetic hot-plasma dielectric tensor and select not only $k_\parallel$ but also $k_x$ to include a first approximation of the power deposition due to harmonic heating. This addition to the basic cold plasma dielectric tensor has not been compared to a kinetic code yet, and does not have the pretension to replace such code. However, a comparison of the power deposition between HFSS and a kinetic code would represent an opportunity to characterize the validity of such a simple approximation for several key ICRH edge parameters and could be the aim of another paper.

\subsection{Ponderomotive forces and sheaths}
Finally, the goal of the present code is to expand the tools readily available for ICRF antenna design. In this objective, our final tool should be able to quickly assess several key features of a future antenna. The antenna coupling resistance, the reflection measured at the antenna circuit and the antenna line voltages are readily available outputs from HFSS. Additionally, the code should assess the plasma wall interactions taking place near and far from the antenna. In this aim,  the sheath taking place on plasma facing components and the ponderomotive force taking place near the antenna should be assessed.

The next step foreseen to include ponderomotive effects in the code is the inclusion of these effects in an approximate way following the POND code approach\cite{meneghini2011modeling}. Indeed, the approach used in this code should be easily reproduced for simple geometries with 1D profiles for development purposes and then later extended to 3D profiles.

For the sheath, a first approach to implement the sheath boundary could be to use the sheath equivalent dielectric layer method presented by \textit{Beers et al.} \cite{beers2021}. This method would have the advantage of being easy to implement in HFSS but requires iterations that could be cumbersome to perform for very large and complex 3D models. Furthermore, the sheath obtained will only be valid for one current distribution over the antenna. Another approach could be to use the post-processing method developed in the work \textit{Myra et al.} \cite{Myra_2019} to assess the sheath voltages that can be expected on complex antenna geometries. This method would have the advantage of being fast and could be integrated in an optimization process looking at various electric field distribution. However, the method is not fully self-consistant and can be inaccurate for complex geometry cases.

\setcounter{footnote}{0}
\section{3D Plasmas} \label{sec:3D}
Now that the possibility of implementing a 1D plasma profile in HFSS was demonstrated, the possibility to implement a 3D plasma is explored. The model used roughly represents one third of the WEST tokamak vessel (simplified geometry), with the same plasma parameters as in section \ref{subsec:ANTcomparison} and with the tepid dielectric tensor presented in section \ref{subsec:Tepid}. This simulation considers toroidal and poloidal rotations of the cold plasma conductivity tensor necessary to account for the toroidal and poloidal field component of the total magnetic field of the device\footnote{While the model implements the effect of the poloidal magnetic field, this field should only create a minor correction to the toroidal-only case.}. Here the magnetic field ripple of WEST is not taken into account but could be easily added as a future functionality in the code.
\newline

\begin{figure*}
    \centering
    \includegraphics[width=\linewidth]{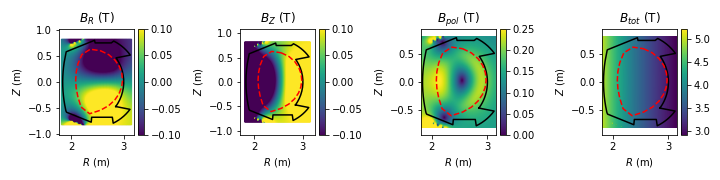}
    \caption{Map of the magnetic fields c   calculated from FreeGS equilibrium.}
    \label{fig:WEST_equilibirum_Bfields}
\end{figure*}

The magnetic equilibrium of the WEST shot \#56898 is recreated with the open source  Free boundary Grad-Shafranov solver for tokamak plasma equilibria (FreeGS) package\footnote{$>>$ pip install freegs}, which provides the magnetic flux surface and the poloidal and toroidal magnetic fields at the time frame of interest \cite{FreeGS,jeon2015development}. The magnetic fields reconstructed using FreeGS are presented in figure \ref{fig:WEST_equilibirum_Bfields}. The figure also displays a clear asymmetric equilibrium, which tends to be the case for most of the WEST shots analyzed. Reaching more symmetric equilibrium could represent an opportunity for WEST to increase the antenna radiation resistance and might lead to a sheath reduction.

\begin{figure*}
    \centering
    \includegraphics[width=\linewidth]{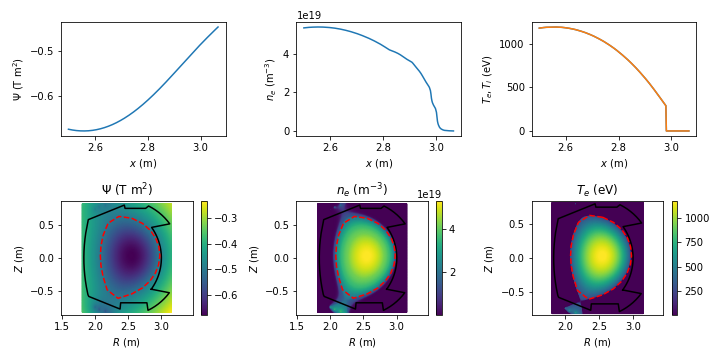}
    \caption{Map of the 1D density and temperature profiles over the magnetic flux surface to create 2D maps.}
    \label{fig:equilibrium_map}
\end{figure*}

The 1D density and temperature profiles are then mapped over magnetic flux surfaces of the equilibrium to create 2D maps as presented in figure \ref{fig:equilibrium_map}. For this first example, the plasma volume implemented in HFSS is restrained to have the simplified equilibrium shape:
\begin{align}
\begin{split}
    R &= R_0 + r \cos(\theta + \arcsin(\delta) \sin(\theta)), \\
    Z &= Z_0 + \kappa r \sin(\theta + \xi\sin(2\theta)).
\end{split}
\end{align}
where $R_0$ and $Z_0$ are the magnetic axis center coordinates and $\kappa$, $\delta$ and $\zeta$ are the ellipticity, triangularity and squareness of the plasma equilibrium computed by FreeGS. This volume is then cut at its first contact with the ICRH antenna limiter. Note that the plasma could also be defined in the full machine volume but needs further work as it might lead to numerical problems linked to slow wave propagation at low densities.

The 2D plasma equilibrium defined in Python is then extended into a 3D map and used to define the final 3D conductivity tensor that is used inside HFSS. The model computes an S-matrix which can be used to compare the simulated antenna characteristics with the experimental values.
\newline

Thus one can compare the antenna coupling resistance and the antenna voltages on each strap measured in the WEST shot \#56898 at 6s with the HFSS modeled antenna ones. The Python package west-ic-antenna
\footnote{$>>$ pip install west-ic-antenna} was used to model the antenna system \cite{Hillairet2020}. This code models the WEST ICRH Antenna by simulating its electrical circuit, with individual antenna components calculated separately. The code uses the antenna S-matrix modeled in HFSS as an input and automatically optimize the tuning capacitors to obtain an ideal match. From the matching capacitor layout the code can calculate several antenna characteristics based on the antenna's power and phase excitation. 

\begin{figure*}
    \centering
    \includegraphics[width=\linewidth]{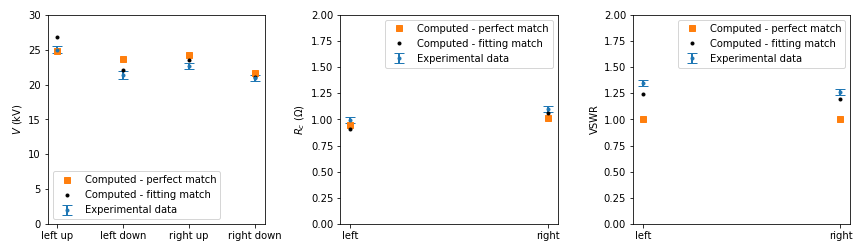}
    \caption{(a) Voltage on straps, (b) coupling resistance and (c) VSWR antenna characteristics obtained experimentally along with the ones computed with the HFSS S-matrix for a perfect match and for a match fitting as much as possible the antenna experimental characteristics provided an maximum error of 0.5 pF imposed to the capacitor layout of the perfectly matched case.}
    \label{fig:expVSmodel_values}
\end{figure*}

With this package one can find the ideal match for the S-matrix computed in HFSS. The voltage, the coupling resistance and the VSWR obtained experimentally are presented in figure \ref{fig:expVSmodel_values} along with an ideal matched case. 

However, the ideal match does not necessarily coincides with the experimental data, besides unavoidable differences between idealized CAD geometries and real geometries. This is seen in the experimental values of the Voltage Standing Wave Ratio (VSWR) presented in figure \ref{fig:expVSmodel_values} (c). To reproduce the experimentally observed reflections, a random uncertainty of 0.5 pF was imposed to the capacitor values obtained for a perfect match. Using those, one can find a new capacitor layout which fits better the experimental data. This is also presented in figure \ref{fig:expVSmodel_values}.

While the experimental and modeled antenna characteristics are already satisfyingly close to each others, one could go further in the code validation by directly measuring the S-matrix of the antenna as was already performed in JET and LHD \cite{Monakhov_2018,DDu2023} and in WEST for lower hybrid current drive \cite{preynas_coupling_2011}. This activity is proposed for the next WEST campaigns.

\section{Conclusion}
In this paper, the possibility to use HFSS as a versatile tool for ICRF simulations was demonstrated. The code was first tested for fast wave propagation inside a uniform cold plasma case. Subsequently, the code was benchmarked with ANTITER for a WEST 1D density profile. The possibility to include a first temperature correction inside the cold plasma model was added. The possibility to treat the slow wave and the LH resonance was tested in the limit of high collisionality. Finally, the code was tested for a 3D WEST plasma taking a third of the WEST tokamak into account. The code reproduces well the results obtained experimentally, further validating it. 

Several possibilities of activity using our new tool were discussed:
\begin{itemize}
    \item to study the edge fields excited by ICRH antennas with complex geometries,
    \item to estimate key antenna parameters like the antenna resistance, the strap voltages, the VSWR, etc;
    \item to approximate core absorption mechanisms,
    \item to include the slow wave dynamics under high collisionality assumptions.
\end{itemize}

Several future amelioration, like the study of sheath and the study of ponderomotive force using the new HFSS tool are under discussions. The study of slow wave propagation could also be pushed further for more complex antenna geometries and for physical collision values. 
\newline

The new tool based on HFSS here presented will be used extensively for the final Travelling Wave Array (TWA) design foreseen for WEST. The same tool is also under study to be used for multipactor analysis in WEST.

\ack \

The authors would like to thank Dr. Laurent Colas for sharing his knowledge on PMLs.

The authors would like to thank Dr. Ben Dudson for his open source code freegs.

The authors would like to thank Dr. Tom Wauters for our shared discussions on neutral collisionality.
\newline

This work has been carried out within the framework of the EUROfusion Consortium, funded by the European Union via the Euratom Research and Training Programme (Grant Agreement No 101052200 — EUROfusion). Views and opinions expressed are however those of the author(s) only and do not necessarily reflect those of the European Union or the European Commission. Neither the European Union nor the European Commission can be held responsible for them.

\section*{References}
\bibliographystyle{iopart-num}%
\bibliography{main}%

\end{document}